\documentclass[twocolumn,aps,showpacs,amssymb,floatfix]{revtex4}
\newcommand{\be}{\begin{equation}}
\newcommand{\ee}{\end{equation}}
\newcommand{\beq}{\begin{eqnarray}}
\newcommand{\eeq}{\end{eqnarray}}
\usepackage{bm}
\usepackage{graphicx}%
\usepackage{epsf}
\usepackage{epsfig}
 
 \usepackage{amsmath}
\usepackage{amsfonts,amssymb}

\usepackage{dsfont}
\usepackage{slashed}
\usepackage{booktabs}

\begin{document}

   \title{Axial Nucleon form factors from lattice QCD}


\author{C.~Alexandrou~$^{(a,b)}$, M. Brinet~$^{(c)}$, J.~Carbonell~$^{(c)}$, M. Constantinou~$^{(a)}$,  P.~A.~Harraud~$^{(c)}$,
P.~Guichon~$^{(d)}$, 
K.~Jansen~$^{e}$, T.~Korzec~$^{(a,f)}$, M.~Papinutto~$^{(c)}$ }
\affiliation{$^{(a)}$ Department of Physics, University of Cyprus, P.O. Box 20537, 1678 Nicosia, Cyprus\\
 $^{(b)}$ Computation-based Science and Technology Research
    Center, Cyprus Institute, 20 Kavafi Str., Nicosia 2121, Cyprus \\
$^{(c)}$ Laboratoire de Physique Subatomique et Cosmologie,
               UJF/CNRS/IN2P3, 53 avenue des Martyrs, 38026 Grenoble, France\\
$^{(d)}$ CEA-Saclay, IRFU/Service de Physique Nucl\'eaire, 91191 Gif-sur-Yvette, France\\
$^{(e)}$ NIC, DESY, Platanenallee 6, D-15738 Zeuthen, Germany\\
         \vspace{0.2cm}
 $^{(f)}$ Institut f\"ur Physik
   Humboldt Universit\"at zu Berlin, Newtonstrasse 15, 12489 Berlin, Germany}

 \begin{abstract}

We present results on the  nucleon axial  form factors within
lattice QCD using two flavors of degenerate twisted mass
fermions. Volume effects are 
 examined using simulations at two volumes of spatial length $L=2.1$~fm
and $L=2.8$~fm. 
Cut-off effects  are investigated using three different
values of the lattice spacings, namely $a=0.089$~fm, $a=0.070$~fm and $a=0.056$~fm.
The nucleon axial charge  is obtained in the
continuum limit and chirally extrapolated to
the physical pion mass 
enabling comparison with experiment.

 \end{abstract}

\begin{minipage}{0.5\linewidth}
\includegraphics[width=0.7\linewidth]{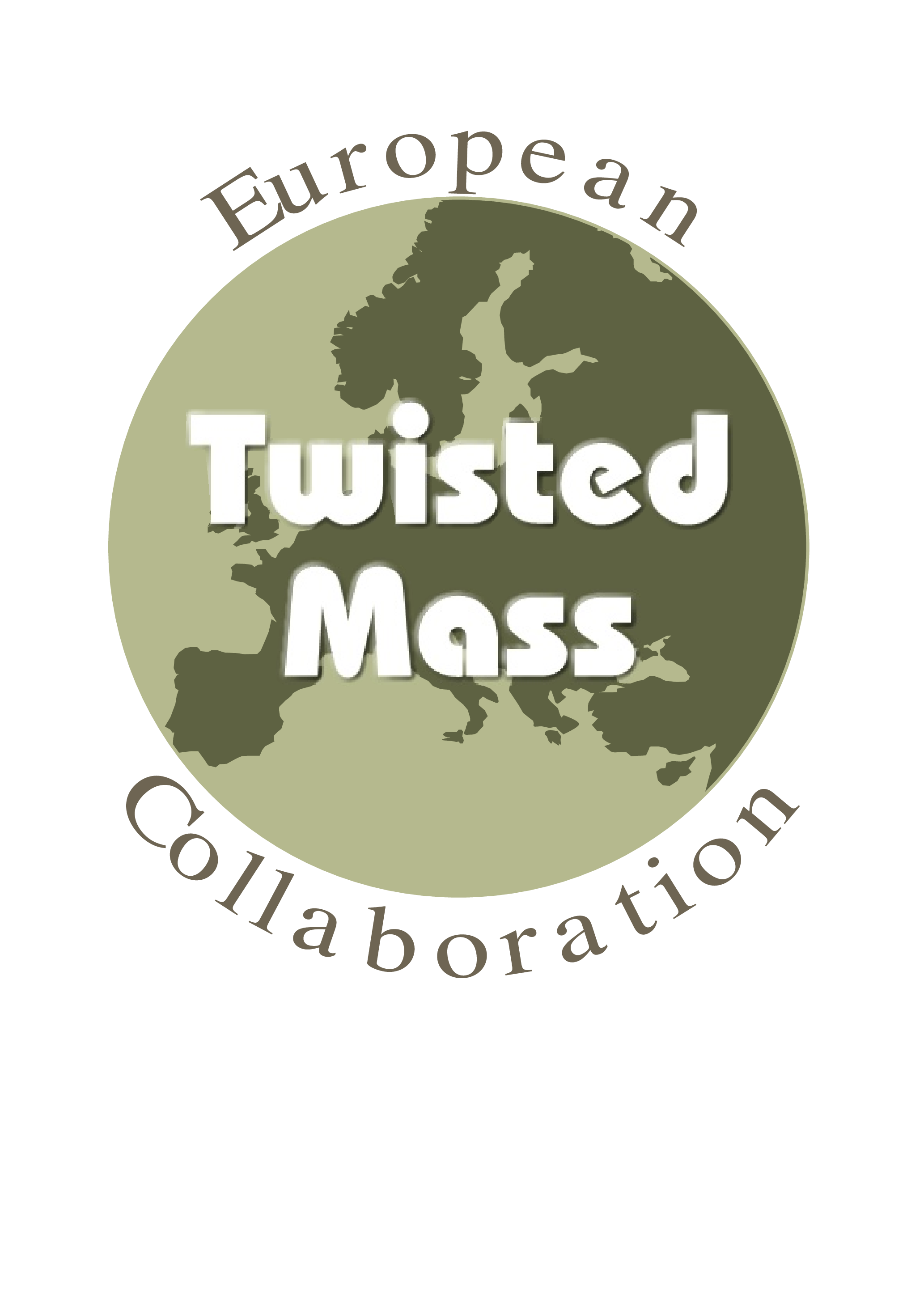}
\end{minipage}\hfill

\maketitle

\setcounter{figure}{\arabic{figure}}

\newcommand{\twopt}[5]{\langle G_{#1}^{#2}(#3;\mathbf{#4};\Gamma_{#5})\rangle}
\newcommand{\threept}[7]{\langle G_{#1}^{#2}(#3,#4;\mathbf{#5},\mathbf{#6};\Gamm
a_{#7})\rangle}

\newcommand{\Op}{\mathcal{O}} 
\newcommand{\C}{\mathcal{C}} 
\newcommand{\eins}{\mathds{1}} 

\bibliographystyle{apsrev}                     

\section{Introduction}
The nucleon (N)  form factors  are fundamental hadronic observables
that probe the structure of the nucleon. Experiments to measure the electromagnetic nucleon form factors
have been carried out since the 50's.
A new generation
of experiments using polarized beams and targets
are currently under way at major facilities in order
to measure the nucleon form factors more accurately 
and  at higher values of the momentum transfer.
The nucleon form factors connected to the
axial vector current
 are more difficult to measure and
therefore less accurately known
than its electromagnetic form factors.  As in the electromagnetic case
the nucleon matrix element of the
axial vector current is written
in terms of two Lorenz invariant form factors, the axial form factor
$G_A(q^2)$   and the induced
pseudo-scalar form factor,
$G_p(q^2)$ where $q^2$ is the momentum transfer squared.
The nucleon axial charge  $g_A=G_A(0)$,
which  can be determined from $\beta-$decay, is known to a high precision.
The $q^2$-dependence of  $G_A(q^2)$  has been studied
from neutrino scattering~\cite{Ahrens:1988rr} 
and pion electroproduction~\cite{Bernard:1992ys,Choi:1993vt} processes. 
The  nucleon induced pseudo-scalar
form factor, $G_p(q^2)$, is even less well known.  Muon capture at low $q^2$ values~\cite{Gorringe:2002xx} and  pion electroproduction
for larger $Q^2$~\cite{Bernard:1992ys,Choi:1993vt}  are the main experimental sources
 of information for $G_p(Q^2)$.
Both $G_A(q^2)$ and $G_p(q^2)$ have been discussed
within chiral effective theories~\cite{Bernard:2001rs,Schindler:2006it}.
In this work we present results on these form factors obtained in lattice QCD using two degenerate
 light quarks  ($N_F{=}2$) in the twisted mass formulation~\cite{Shindler:2007vp}.

Twisted mass fermions~\cite{Frezzotti:2000nk}  provide an attractive  formulation of lattice QCD that
allows for automatic ${\cal O}(a)$ improvement, infrared regularization 
of small
eigenvalues and fast dynamical 
simulations. For the
calculation of the nucleon form factors in which we are interested in this work,the automatic  
 ${\cal O}(a)$ improvement is particularly relevant 
since it is achieved  by tuning only one parameter in the action,
requiring no further improvements on the operator level.

The action for two degenerate flavors of quarks
 in twisted mass QCD is given by
   \begin{equation}
      S=S_g + a^4 \sum_x \bar\chi(x) \left[D_W
 {+} m_{\rm crit}
 {+} i\gamma_5\tau^3\mu \right]\chi(x)\,,
   \end{equation}
where $D_W$ is the Wilson Dirac operator and we use the  tree-level Symanzik improved
gauge action $S_g$~\cite{Weisz:1982zw}. The quark fields $\chi$
are in the so-called ``twisted basis'' 
obtained from the ``physical basis''
at
maximal twist by a simple transformation:
\be
\psi {=} \frac{1}{\sqrt{2}}[{\bf 1} + i\tau^3\gamma_5]\chi \quad {\rm and}
\quad \bar\psi {=} \bar\chi \frac{1}{\sqrt{2}}[{\bf 1} + i\tau^3\gamma_5]\,.
\ee
We note that, in the continuum, this action is equivalent to the standard
QCD action. A
crucial advantage is the fact that by tuning a single parameter,
namely the bare untwisted quark mass to its critical value $m_{\rm
  cr}$, 
a wide class of physical observables are automatically ${\cal O}(a)$
improved~\cite{Shindler:2007vp}. A disadvantage is  the explicit flavor symmetry breaking. In
a recent paper we have checked that this breaking is small for the baryon
observables under consideration in this work and for the lattice spacings
that we use~\cite{Alexandrou:2009xk,Alexandrou:2009qu,Drach:2009dh,Alexandrou:2008tn,Alexandrou:2007qq}. To
extract the nucleon FFs we need to evaluate the nucleon matrix
elements $\langle N(p^\prime,s^\prime) | {A_\mu^a} | N(p,s) \rangle$, where
$|N(p^\prime,s^\prime)\rangle$, $|N(p,s)\rangle$ are nucleon states with
final (initial) momentum $p^\prime (p)$ and spin $s^\prime (s)$. Due to its isovector nature, the axial
vector current, defined by
\be A_\mu^a(x) {=} \bar\psi(x) \gamma_\mu \gamma_5
\frac{\tau^a}{2} \psi(x)\,,
\ee
receives contributions only from the connected 
diagram for $a=1,\,2$ and up to ${\cal O}(a)$ for $a=3$.
Simulations including a dynamical strange quark are also available within the
twisted mass formulation. Comparison of the nucleon mass obtained with two
dynamical flavors and the nucleon mass including a dynamical strange quark
has shown negligible dependence on the dynamical strange quark~\cite{Drach:2010}. 
We therefore expect the results on the nucleon form factors to show little
sensitivity on
a dynamical strange quark as well.

  The axial current matrix element of the nucleon  $ \langle N(p^\prime,s^\prime) | A^a_\mu(0) | N(p,s) \rangle$ can be expressed in terms of the form factors
  $ G_A$ and $G_p$ as
\beq 
\langle N(p',s')|A_\mu^3|N(p,s)\rangle= i \Bigg(\frac{
            m_N^2}{E_N({\bf p}')E_N({\bf p})}\Bigg)^{1/2} \nonumber \\
            \bar{u}_N(p',s') \Bigg[
            G_A(q^2)\gamma_\mu\gamma_5 
            +\frac{q_\mu \gamma_5}{2m_N}G_p(q^2) \Bigg]\frac{1}{2}u_N(p,s).
\label{axial ff}
\eeq 

In this work we consider simulations at three values of the coupling constant
spanning lattice spacings from about 0.05~fm to 0.09~fm. This enables us to
obtain results in the continuum limit. We find that cut-off effects are
small for this range of lattice spacings. We also examine finite size effects
on the axial form factors by comparing results on two lattices of spatial length $L=2.1$~fm and $L=2.8$~fm~\cite{Alexandrou:2010,Alexandrou:2009ng,Alexandrou:2008rp}. 

\section{Lattice evaluation}

\subsection{Correlation functions} 

 The nucleon interpolating field in the physical basis is given by
\be
J(x) = \epsilon^{abc} \left[u^{a \top}(x) \C\gamma_5 d^b(x)\right] u^c(x)
\ee
 and can be written in the twisted basis  at maximal twist as
\be
\tilde{J}(x) {=} {\frac{1}{\sqrt{2}}[\eins + i\gamma_5]}\epsilon^{abc} \left[ {\tilde{u}}^{a \top}(x) \C\gamma_5 \tilde{d}^b(x)\right] {\tilde{u}}^c(x).
\ee
The transformation of the axial vector current, $A_\mu^a(x)$, to the
twisted basis leaves the form of 
$A_\mu^{3}(x)$ 
unchanged. The axial
renormalization constant $Z_A$  is determined non-perturbatively in
the RI'-MOM scheme using two
approaches~\cite{Dimopoulos:2007qy,Constantinou:2010gr} and
~\cite{Alexandrou:2010me,Constantinou:2010dn} both of which yield
consistent values. We use the values of $Z_A$ found in latter
approach~\cite{Alexandrou:2010me}, which employs a momentum
source~\cite{Gockeler:1998ye} and  a perturbative subtraction of
${\cal O}(a^2)$ terms~\cite{Martha, Alexandrou:GPDs}. This subtracts
the leading cut-off effects yielding  only a very weak dependence of
$Z_A$ on $(ap)^2$ for which the $(ap)^2\rightarrow 0$ limit can be
reliably taken. It was also shown with high accuracy that the quark
mass dependence of $Z_A$ is negligible. We find the values
\be
Z_A{=}0.757(3)\,,\,\, 0.776(3)\,,\,\,0.789(3)
\ee
at $\beta{=}$3.9, 4.05 and 4.2
respectively. These are the values of $Z_A$ which we use in this
work to renormalize the lattice matrix element.

\begin{figure}
 \includegraphics[width=\linewidth]{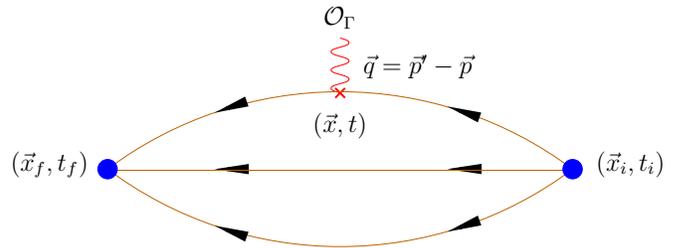}
\caption{Connected nucleon three-point function.}
\label{fig:connected_diagram}
\end{figure}

In order to increase the overlap with the proton state and
decrease overlap with excited states we use Gaussian smeared quark
fields~\cite{Alexandrou:1992ti,Gusken:1989} for the construction of
the interpolating fields:
\beq
q_{\rm smear}^a(t,\vec x) &=& \sum_{\vec y} F^{ab}(\vec x,\vec y;U(t))\ q^b(t,\vec y)\,,\\
F &=& (\eins + {\alpha} H)^{n} \,, \nonumber\\
H(\vec x,\vec y; U(t)) &=& \sum_{i=1}^3[U_i(x) \delta_{x,y-\hat\imath} + U_i^\dagger(x-\hat\imath) \delta_{x,y+\hat\imath}]\,. \nonumber
\eeq
In addition, we apply APE-smearing to the gauge fields $U_\mu$ entering 
the hopping matrix $H$.
The smearing parameters   are the same as those used for our calculation of baryon masses
with $\alpha$ and $n$ optimized for the nucleon
ground state~\cite{Alexandrou:2008tn}. The values are: $\alpha=4.0$ and $n=50$, $70$ and $90$ 
for $\beta=3.9$, $4.05$ and $4.2$ respectively.

In order to calculate the  nucleon matrix element of Eq.~(\ref{axial ff}) 
we calculate 
the two-point and three-point functions defined by
\hspace{-0.55cm}
\beq
\hspace{-0.55cm}G(\vec q, t_f)\hspace{-0.15cm}&=&\hspace{-0.25cm}\sum_{\vec x_f} \, e^{-i\vec x_f \cdot \vec q}\, 
     {\Gamma^0_{\beta\alpha}}\, \langle {J_{\alpha}(t_f,\vec x_f)}{\overline{J}_{\beta}(t_i,\vec{x}_i)} \rangle \\
\hspace{-0.5cm}
G^\mu(\Gamma^\nu,\vec q, t) \hspace{-0.15cm}&=&\hspace{-0.25cm}\sum_{\vec x, \vec x_f} \, e^{i\vec x
  \cdot \vec q}\,  \Gamma^\nu_{\beta\alpha}\, \langle
{J_{\alpha}(t_f,\vec x_f)} A^\mu(t,\vec x) {\overline{J}_{\beta}(t_i,\vec{x}_i)}\rangle,\nonumber
\eeq
where ${\Gamma^0}$ and ${\Gamma^k}$ are the projection matrices:
\be
{\Gamma^0} = \frac{1}{4}(\eins + \gamma_0)\,,\quad {\Gamma^k} =
i{\Gamma^0} \gamma_5 \gamma_k\,.
\ee
The kinematical setup that we used is illustrated in
Fig.~\ref{fig:connected_diagram}: We create the nucleon at
$t_i{=}0$, at $\vec x_i {=} 0$ (source) and annihilate it at a later time
$t_f$ with $\vec p^\prime {=} 0$ (sink). The current couples to
a quark at an intermediate time $t$.
For our kinematics $\vec{q}=-\vec{p}$. 
From now on all quantities are given in Euclidean space and
where $Q^2=-q^2$ is the Euclidean momentum transfer squared.
The  leading time dependence
of the Euclidean time evolution
and the overlap factors cancel for the ratio
\be
R^{\mu}(\Gamma,\vec q,t)= \frac{G^\mu(\Gamma,\vec q,t) }{G(\vec 0,
  t_f)}\ \sqrt{\frac{G(\vec p_i, t_f-t)G(\vec 0,  t)G(\vec0,
    t_f)}{G(\vec 0  , t_f-t)G(\vec p_i,t)G(\vec p_i,t_f)}}\,,
\label{ratio}
\ee
 yielding a time-independent value 
\be
\lim_{t_f-t\rightarrow \infty}\lim_{t-t_i\rightarrow \infty}R^{\mu}(\Gamma,\vec q,t)=\Pi^\mu (\Gamma,\vec q) \,.
\label{plateau}
\ee
We refer to the range of $t$-values where this asymptotic behavior is observed
within our statistical precision as the plateau range.
\begin{figure}\vspace*{-4cm}
 \includegraphics[width=1.05\linewidth, height=1.8\linewidth]{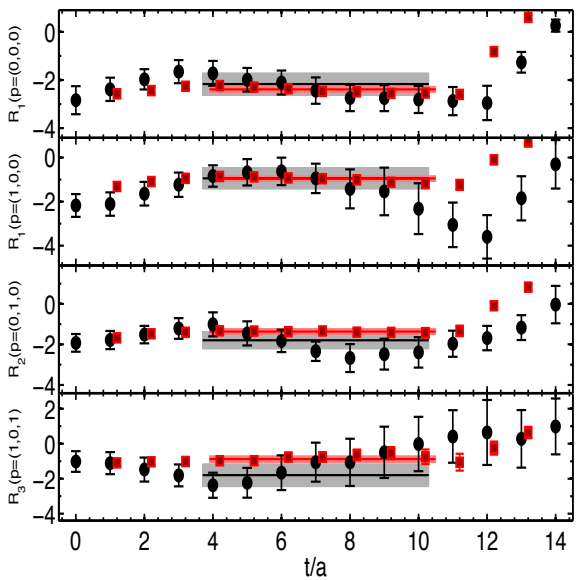}
\vspace*{-4cm}
\caption{The ratio of  Eq.~(\ref{ratio}) for 
representative momentum combinations at $\beta=3.9$ and different values
of $\mu$. The filled (black) circles  show results
with a sink-source separation $t_f/a=14$ and the filled (red) squares for $t_f/a=12$,  shifted to the left by one time-slice.
}
\label{fig:plateau}
\end{figure}
As mentioned already, only 
the connected diagram contributes. It is  calculated
 by performing sequential inversions through the sink yielding the
form factors at all possible
momentum transfers and current orientation $\mu$.
Since we use sequential inversions through the sink we need to fix the
sink-source separation. Statistical errors increase rapidly as we increase
the sink-source separation. Therefore we need to choose the smallest 
possible that still ensures that the nucleon ground state dominates
when measurements are made at different values of $t$.
 In order to check that a sink-source 
time separation of $\sim 1$~fm is sufficient for the isolation of the 
nucleon ground state we compare the results at $\beta{=}3.9$ obtained with $t_f/a{=}12$ i.e. $t_f\sim 1$~fm
with those obtained when we increase to
 $t_f/a{=}14$~\cite{Alexandrou:2008rp}. 
As can be seen
in Fig.~\ref{fig:plateau}, where we plot $R^\mu(\Gamma,\vec{q},t)$, 
the plateau values are compatible 
yielding  the
same plateau value for the two time separations. This 
means  that the shorter sink-source separation is sufficient and the
 ground state of the nucleon dominates in the plateau region. We therefore use in all of
our analysis   $t_f-t_i\sim 1$fm.

New inversions are necessary every time a different choice of the projection matrix $\Gamma^\alpha$ is made. In this work, we consider  choices, which
are optimal for the form factors considered. Namely we use the spatial
$\Gamma$'s and consider the spatial component of the current i.e. 
we extract the form factors from
\be
\Pi^{i}(\Gamma^k,\vec q){=}\frac{ic}{4m}\left[\frac{q_k q_i}{2m}\ G_p(Q^2){-}(E{+}m)\delta_{i,k}\ G_A(Q^2)\right],
\label{plateau1}
\ee
 where $k=1,2,3, \quad{\rm and}\quad c=\sqrt{\frac{2m^2}{E(E+m)}}$.

\subsection{Simulation details}

The input parameters of the calculation, namely $\beta$, $L/a$ and $a\mu$ 
are summarized in Table~\ref{Table:params}. The  lattice spacing $a$ 
is taken from the nucleon mass as described in the next section.
The pion mass values, spanning a mass range 
from 260 to 470~MeV, are taken 
from Ref.~\cite{Urbach:2007}.  
At $m_{\pi}\approx 300$ MeV and $\beta{=}3.9$ we have simulations 
for lattices of spatial size $L{=}2.1$~fm and $L{=}2.8$~fm  
allowing to investigate finite size effects. 
Finite lattice spacing effects are studied using three sets of 
results at $\beta{=}3.9$, $\beta{=}4.05$ and $\beta{=}4.2$ for the lowest
and largest pion mass available in this work.
These sets of gauge ensembles allow us to estimate all the systematic
errors in order to
produce reliable predictions for the nucleon axial form factors.
\begin{widetext}
\begin{center}
\begin{table}[h]
\begin{tabular}{c|llllll}
\hline\hline
\multicolumn{6}{c}{$\beta=3.9$, $a=0.089(1)(5)$~fm,   ${r_0/a}=5.22(2)$}\\\hline 
$24^3\times 48$, $L=2.1$~fm &$a\mu$         &   &    0.0040      &   0.0064     &  0.0085     &   0.010 \\ 
                               & Stat. &  &943 &553 & 365 &477 \\ 
                               &$m_\pi$~(GeV) &  & 0.3032(16) & 0.3770(9) & 0.4319(12) & 0.4675(12)\\
                               &$m_\pi L$     &    & 3.27       & 4.06      & 4.66       & 5.04     \\
$32^3\times 64$, $L=2.8$~fm  &$a\mu$ & 0.003 & 0.004 & & & \\
                               & Stat. & 667  &351 & & & \\
                               & $m_\pi$~(GeV)& 0.2600(9)   & 0.2978(6) & & &  \\
                               & $m_\pi L$    & 3.74        & 4.28      &&& \\\hline \hline
\multicolumn{6}{c}{ $\beta=4.05$, $a=0.070(1)(4)$~fm, ${r_0/a}=6.61(3)$ }\\
\hline
$32^3\times 64$, $L=2.13$~fm &$a\mu$         & 0.0030     & 0.0060     & 0.0080     & \\
                               & Stat.   &447 &326 &  419 &\\
                               &$m_\pi$~(GeV) & 0.2925(18) & 0.4035(18) & 0.4653(15) &  \\
                               &$m_\pi L$     & 3.32       &   4.58     & 5.28       &       \\ \hline\hline
\multicolumn{6}{c}{ $\beta=4.2$, $a=0.056(1)(4)$~fm  ${r_0/a}=8.31$}\\\hline
$32^3\times 64$, $L=2.39$~fm &$a\mu$          & 0.0065     &     & \\
                               & Stat.       & 357 &  & & \\
                               &$m_\pi$~(GeV) & 0.4698(18) &  &  \\
                               &$m_\pi L$     & 4.24       &      & \\
$48^3\times 96$, $L=2.39$~fm &$a\mu$         & 0.002      &      &     & \\
                               & Stat.      & 245          &  &  & & \\
                               &$m_\pi$~(GeV) & 0.2622(11) &  &  &  \\
                               &$m_\pi L$     & 3.55       &  &      &       \\ \hline

\end{tabular}
\caption{Input parameters ($\beta,L,a\mu$) of our lattice calculation and corresponding lattice spacing ($a$) and pion mass ($m_{\pi}$).}
\label{Table:params}
\vspace*{-.0cm}
\end{table} 
\end{center}

\end{widetext}

\subsection{Determination of the lattice spacing}
\begin{figure}
 \includegraphics[width=\linewidth]{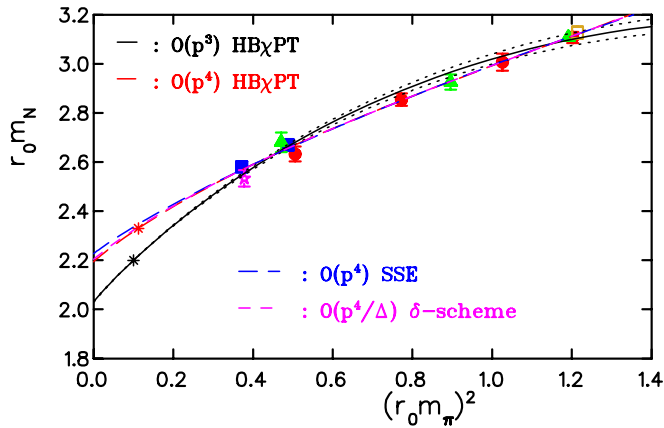}
\caption{ Nucleon mass in units of $r_0$ at three lattice spacings and spatial
lattice size $L$ such that  $m_\pi L> 3.3$. The solid (black) and dashed (red) lines
are  fits to ${\cal O}(p^3)$ and ${\cal O}(p^4)$ HB$\chi$PT. The physical point is shown with the asterisks. Results at $\beta=3.9$ and $24^3\times 48$ are 
shown with filled (red) circles, at $\beta=3.9$ and $32^3\times 64$ with the  filled (blue) squares, at $\beta=4.05$ and $32^3\times 64$ with the filled (green) triangles, at  $\beta=4.2$ and $32^3\times 64$  with the open (yellow) square
and at  $\beta=4.2$ and $48^3\times 96$  with the star (magenta).}
\label{fig:nucleon scale}
\end{figure}

The nucleon mass has been computed on the same
ensembles that are now
used here for the computation of the nucleon axial
form factors~\cite{Alexandrou:2008tn}. Therefore we can use the nucleon mass at the physical point
to set the scale. We show in
Fig.~\ref{fig:nucleon scale} results at three values of the lattice
spacings corresponding to $\beta{=}3.9$, $\beta{=}4.05$ and
$\beta{=}4.2$. As can be seen, cut-off effects are negligible and we can
therefore use continuum chiral perturbation theory to extrapolate to
the physical point. For the observables discussed in this work the
nucleon mass at the physical point is the most appropriate quantity  to
set the scale. This also provides a cross-check for the determination
of the lattice spacing as compared to the pion decay constant used in
the meson sector. If lattice artifacts are under control then these
two determinations should be consistent, under the 
assumption that quenching effects
due to the absence of the strange quark from the sea are small for
these quantities.
In order to correct  for
volume effects we use  chiral perturbation theory to take into
account volume corrections coming from pions propagating around the
lattice, following Ref.~\cite{AliKhan:2003cu}. A similar analysis
was carried out in Ref.~\cite{Alexandrou:2009qu} at $\beta=3.9$ and $\beta=4.05$
and we refer
to this publication for additional details. 
In addition, this work is extended by an analysis of
results at $\beta=4.2$. In Table~\ref{Table:mN} we give the volume
corrected nucleon mass.
\begin{table}[h]
\begin{center}
\begin{tabular}{c|c|c|c}
\hline\hline
$am_\pi$ & $Lm_\pi$ & $am_N$ & $am_N(L\rightarrow \infty)$ \\\hline 
\multicolumn{4}{c}{ $\beta=3.9$} \\\hline
0.2100(5) &5.04 & 0.5973(43) & 0.5952 \\
0.1940(5) &4.66 & 0.5786(67) & 0.5760 \\
0.1684(2) &4.06 & 0.5514(49) & 0.5468\\
0.1362(7) &3.27 & 0.5111(58) & 0.5043\\
0.1338(2) &4.28 & 0.5126(46) & 0.5115\\
0.1168(3) &3.74 & 0.4958(43) & 0.4944\\
\multicolumn{4}{c} {$\beta=4.05$} \\\hline
0.1651(5) & 5.28 & 0.4714(31) &  0.4702\\
0.1432(6) & 4.58 & 0.4444(47) &  0.4426\\
0.1038(6) & 3.32 & 0.4091(60) &  0.4056\\
\multicolumn{4}{c}{ $\beta=4.2$} \\\hline
0.1326(5) & 4.24 & 0.380(3)& 0.3763 \\
0.0740(3) & 3.55 & 0.306(4)& 0.3049\\
\hline
\end{tabular}
\caption{Results on the nucleon mass. The last column gives
  the values after a volume correction.}
\label{Table:mN}
\end{center}
\end{table}

To chirally extrapolate we use the well-established ${\cal O}(p^3)$ result of
heavy baryon chiral perturbation theory (HB$\chi$PT) given by
\be  m_N = {m_N^0}-{4c_1}m_\pi^2 -\frac{3 g_A^2 }{16\pi f_\pi^2} m_\pi^3 .\ee
We perform a fit  to the volume corrected results at $\beta{=}3.9$, $\beta{=}4.05$ and $\beta{=}4.2$  and extract  $r_0{=}0.462(5)$~fm.
 Fitting instead to
the  $\beta{=}3.9$ and $\beta{=}4.05$ results we find  $r_0{=}0.465(6)$~fm showing
that indeed cut-off effects are small. To estimate
the error due to the chiral extrapolation we use HB$\chi$PT to ${\cal O}(p^4)$, which leads to
$r_0{=}0.489(11)$.
 We take the difference between the ${\cal O}(p^3)$
and  ${\cal O}(p^4)$ mean values 
as an estimate of the uncertainty due to the chiral extrapolation.
 Fits to other higher order $\chi$PT formulae are also shown
 in Fig.~\ref{fig:nucleon scale}. These are 
described in Ref.~\cite{Alexandrou:2008tn} and 
are consistent with  ${\cal O}(p^4)$ HB$\chi$PT.
Using $r_0{=}0.462(5)(27)$ and  the computed $r_0/a$ ratios we obtain
\beq
a_{\beta=3.9}&=&0.089(1)(5)\,, \nonumber\\
a_{\beta=4.05}&=&0.070(1)(4)\,,\nonumber\\
a_{\beta=4.2}&=&0.056(2)(3)\,.\nonumber
\eeq
 These values are consistent with the lattice spacings determined from
$f_\pi$ and will be used
for converting to physical units in what follows.
 We note that results on the nucleon mass using twisted mass fermions agree with
those obtained using other lattice ${\cal O}(a^2)$ formulations
for lattice spacings below 0.1~fm~\cite{Alexandrou:2009qu}.

\section{Results}
In the first subsection we discuss our results on the nucleon axial charge and
in the second subsection we discuss the momentum dependence of the 
axial $G_A(Q^2)$ and the induced pseudo-scalar $G_p(Q^2)$.

\subsection{Axial charge}
\begin{figure}
      \includegraphics[width=\linewidth]{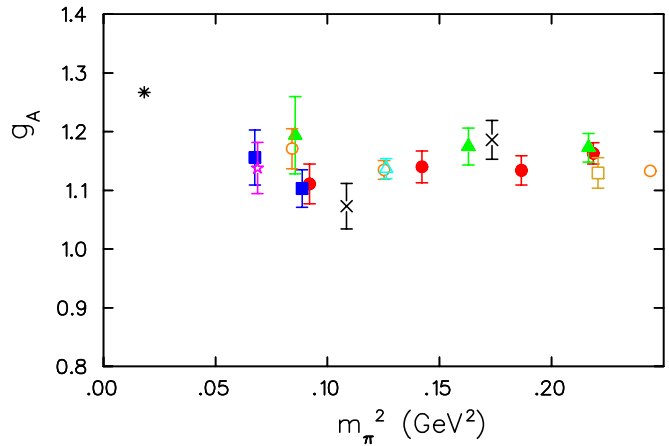}
\caption{The nucleon axial charge. Results using $N_F=2$ twisted mass fermions
are shown using the same notation as that of Fig.~\ref{fig:nucleon scale}. Crosses show results obtained using $N_F=2+1$ DWF,  circles
are results in a mixed action approach on a lattice of size $20^3\times 64$ and
the triangle on a lattice of size $28^3 \times 64$.  }
\label{fig:axial_charge}
   \end{figure}

Our lattice results on the nucleon axial charge are shown in
Fig.~\ref{fig:axial_charge} and listed in
Table~\ref{Table:gA raw}. In the
same figure we also show results obtained using $N_F=2+1$
domain wall fermions (DWF) by the RBC-UKQCD collaborations~\cite{Yamazaki:2009zq} and 
using a mixed action with 2+1 flavors of asqtad sea and domain wall 
valence fermions  by LHPC~\cite{Bratt:2010jn}.
The first observation is that results at our three different 
lattice spacings are within error bars. The second observation is
that results at two different volumes are also consistent. 
The third observation is that  there is agreement among
lattice results using different lattice actions even before taking
the continuum and infinite volume limit.
\begin{table}[h]
\begin{center}
\begin{tabular}{c|c|c|c}
\hline\hline
$m_\pi$ & $Lm_\pi$ & $g_A$ & $g_A(L\rightarrow \infty)$ \\\hline 
\multicolumn{4}{c}{ $\beta=3.9$} \\\hline
    0.4675 & 5.04 &  1.163(18)   &1.167\\
    0.4319 & 4.66 &  1.134(25)   &1.140\\
    0.3770 & 4.06 &  1.140(27)   &1.150\\
    0.3032 & 3.27 &  1.111(34)   &1.133\\
    0.2978 & 4.28 &  1.103(32)   &1.106\\
    0.2600 & 3.74 &  1.156(47)   &1.162\\\hline
\multicolumn{4}{c} {$\beta=4.05$} \\\hline
    0.4653 & 5.28 &  1.173(24)   &1.177\\
    0.4035 & 4.58 &  1.175(31)   &1.182\\
    0.2925 & 3.32 &  1.194(66)   &1.218\\\hline
\multicolumn{4}{c}{ $\beta=4.2$} \\\hline
    0.4698 & 4.24 &   1.130(26)  & 1.144\\
    0.2622 & 3.55 &  1.138(43)   &1.146\\\hline
\end{tabular}
\caption{Results using $N_F=2$ twisted mass fermions (TMF) on the axial nucleon charge. The last column gives
  the values after a volume correction.}
\label{Table:gA raw}
\end{center}
\end{table}

\subsubsection{Finite volume effects}
\begin{figure}
\includegraphics[width=\linewidth]{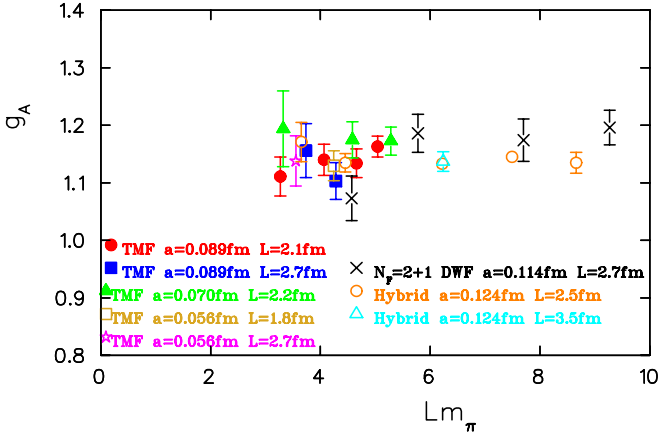}            
\caption{The nucleon axial charge as a function of $Lm_\pi$. The notation
is the same as that of Fig.~\ref{fig:axial_charge}.}
\label{fig:gA_Lmpi}
\end{figure}

In order to assess  volume effects we plot 
in Fig.~\ref{fig:gA_Lmpi}  results on $g_A$ versus $Lm_\pi$. 
Besides TMF results we show the results obtained using $N_F=2+1$ 
DWF~\cite{Yamazaki:2009zq} as well as within the mixed action approach~\cite{Bratt:2010jn}. As can be seen
the results are consistent with a constant in the whole range of $Lm_\pi$
spanned.  In particular we do not observe a decrease in the value of $g_A$
for values of  $Lm_\pi\sim 3.3$. Therefore, given  that finite volume effects are negligible for the smallest value of  $m_\pi L =3.3$ as compared to the value
we find at  $m_\pi L = 4.3$, we conclude that for all of our data 
for which $m_\pi L>3.3$  volume effects are small.
\begin{figure}\vspace*{0.3cm}
\includegraphics[width=\linewidth]{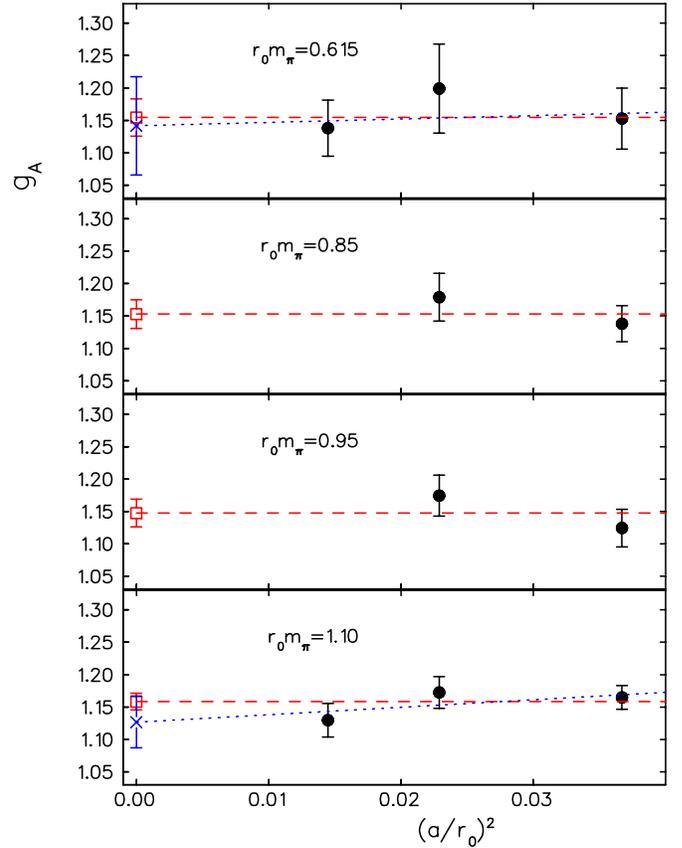}            
\caption{The nucleon axial charge as a function of the
lattice spacing in units of $r_0$ at:  $r_0 m_\pi=0.615$, $r_0 m_\pi=0.85$,  $r_0 m_\pi=0.95$, $r_0 m_\pi=1.10$, from top to bottom. We use $r_0=0.462(5)$
extracted from the nucleon mass. In the upper most and lower most graphs
we show both the linear (dotted line) and constant fits (dashed line).}
\label{fig:gA_continuum}
\end{figure}

We can estimate the volume correction on $g_A$ within
 heavy baryon chiral perturbation theory (HB$\chi$PT) in the so called small scale expansion (SSE)~\cite{Hemmert:1997ye}, which includes
explicitly the $\Delta$ degree of freedom. In this scheme one expands the
results in powers of a small parameter $\epsilon$, which denotes
small pion four-momenta, the pion mass, baryon three-momenta and the nucleon-$\Delta$ mass splitting in the chiral limit. Writing
\be
g_A(m_\pi,\infty)=g_A(m_\pi, L)-\Delta g_A(m_\pi,L)
\ee
 the dependence
of $g_A$ on the spatial length $L$ of the lattice is given by~\cite{Khan:2006de}\beq
&{} &\Delta g_A(m_\pi,L)=-\frac{g_A^0m_\pi^2}{4\pi^2f_\pi^2}\sum_{\vec n}^\prime\frac{K_1(L|\vec n|m_\pi)}{L|\vec n|m_\pi} \nonumber\\
 &+& \frac{(g_A^0)^3m_\pi^2}{6\pi^2 f_\pi^2}\sum_{\vec n}^\prime \left[K_0(L|\vec n|m_\pi)-\frac{K_1(L|\vec n|m_\pi)}{L|\vec n|m_\pi}\right] \nonumber\\
&+& \frac{c_A^2}{\pi^2 f_\pi^2}\left(\frac{25}{81} g_1 -g_A^0\right) \int_0^\infty \, dy\, y\sum_{\vec n}^\prime\Biggl[K_0(L|\vec n|f(m_\pi,y))
\nonumber \\
&{ }&\hspace*{2.5cm}\left.-\frac{L|\vec n|f(m_\pi,y)}{3} K_1(L|\vec n|f(m_\pi,y))\right] \nonumber \\
&+& \frac{8c_A^2g_a^0}{27\pi^2 f_\pi^2} \int_0^\infty \, dy \, y\sum_{\vec n}^\prime \frac{f(m_\pi,y}{\Delta_0}\Biggl[K_0(L|\vec n|f(m_\pi,y))\nonumber \\
&{}& \hspace*{3cm}\left.-\frac{K_1(L|\vec n|f(m_\pi,y))}{L|\vec{n}|f(m_\pi,y)}\right] \nonumber \\
&+& -\frac{4c_A^2 g_A^0}{27\pi f_\pi^2}\frac{m_\pi^3}{\Delta_0}\sum_{\vec n}^\prime \frac{1}{L|\vec{n}|m_\pi} \, e^{-L|\vec{n}|m_\pi} + {\cal O}(\epsilon^4)
\label{gA vol}
\eeq
with $f(m_\pi,y)=\sqrt{m_\pi^2+y^2+2y\Delta_0}$ and $f_\pi$ the pion
decay constant in the chiral limit which we approximate with its physical
value i.e. we take $f_\pi=0.092$~GeV. In the sum $\sum_{\vec n}^\prime$  all vectors $\vec n$ are summed except $\vec n=\vec 0$. In order 
to estimate the volume correction $\Delta g_A$ we take the experimental
value of  the axial charge
in the chiral limit i.e. $g_A^0 \sim g_A^{\rm exp}=1.267$ and the nucleon - $\Delta$ mass splitting in the chiral limit $\Delta_0=0.2711$. For the $\Delta$ axial
coupling constant we use the SU(4) relation $g_1=9g_A^{\rm exp}/5$ and for the 
nucleon to $\Delta$ axial coupling constants $c_A=1.5$. 
The estimated volume corrected  $g_A$  is given in Table~\ref{Table:gA raw}.

In order to assess cut-off effects we use the simulations at three lattice
spacings at the smallest and largest pion mass used in this work. We take 
as reference pion mass the one computed on the finest lattice and
interpolate results at the other two $\beta$-values to these two reference masses.
In Fig.~\ref{fig:gA_continuum} we show the value of $g_A$ at these reference pion
masses computed in units of $r_0$. We perform a fit to
these data using   a linear form $g_A(a)=g_A(0)+c(a/r_0)^2$. The resulting
fit is shown in Fig.~\ref{fig:gA_continuum}. Setting $c=0$ we obtained
the constant line also shown in the figure. As can be seen, for both large
and small pion masses the slope is consistent with zero yielding a value
in the continuum limit in agreement with the constant fit.
Therefore we conclude that finite $a$ effects are negligible and
for the intermediate pion masses we 
obtained the values in the continuum by fitting our data at $\beta=3.9$ and $\beta=4.05$ to a constant.

\begin{widetext}
\begin{center}
\begin{table}[h]
\begin{tabular}{c|c|c|c|c|c|}
\hline\hline
$r_0m_\pi$  & $g_A(\beta=3.9)$ & $g_A(\beta=4.05)$& $g_A(\beta=4.2)$ & $g_A(a\rightarrow 0)$ &  $g_A(L\rightarrow \infty, a\rightarrow 0)$ \\\hline 
 1.1019& 1.165(18) &  1.173(25) &  1.130(26) &  1.159(13) [1.127(40)]&  1.165(13) [1.144(40)] \\
 1.0   & 1.132(25) &  1.172(33) &            &  1.147(20)            &  1.153(20)             \\
 0.95  & 1.125(29) &  1.175(31) &            &  1.148(21)            &  1.155(21)              \\
 0.85  & 1.138(28) &  1.179(37) &            &  1.153(22)            &  1.165(22)              \\
 0.686 & 1.110(39) &  1.194(66) &            &  1.127(34)            &  1.129(34)               \\
 0.615 & 1.153(47) &  1.199(69) &  1.138(43) &  1.154(29) [1.142(76)]&  1.165(29) [1.156(76)]   \\\hline
\end{tabular}
\caption{In the second, third and fourth column we give the interpolated values of $g_A$ 
at the value of $m_\pi r_0$ given in the first column. We used $r_0/a=5.22(2)$, $6.61(3)$ and $8.31(5)$ for $\beta=3.9$, $4.05$ and $4.2$ respectively. In the fifth column we give the value of $g_A$ after extrapolating to $a=0$ using a constant fit. In the parenthesis we give the corresponding values when using a linear fit. In the last
column we  give the continuum value of $g_A$  for the volume-corrected data.}
\label{Table:gA_continuum}
\end{table}
\end{center}
\end{widetext}

\begin{figure}
\includegraphics[width=\linewidth]{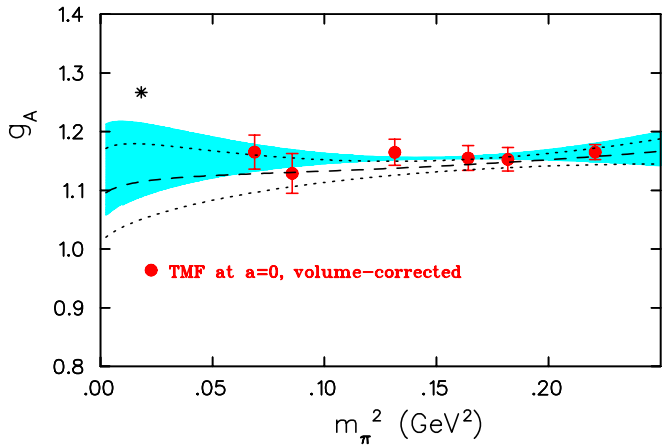}            
\caption{The nucleon axial charge obtained by taking the continuum limit of the volume corrected data. The shaded area shows the best fit to the data
shown on the graph. The dashed line shows the best fit to the raw lattice
data at the three values of $\beta$ with the dotted lines being the associate error band.}
\label{fig:gA_cont}
\end{figure}

The values for $g_A$ found at  six reference pion masses are given in Table~\ref{Table:gA_continuum}. 
We  give both the continuum values obtained using a constant fit 
when no volume corrections are carried out as
well with the volume corrected data.  The volume corrected
data extrapolated to $a=0$ are plotted  in Fig.~\ref{fig:gA_cont}.

\subsubsection{Chiral extrapolation}
Our simulations  cover a
range of pion masses from about 470~MeV down to about
260~MeV. The pion mass dependence for the nucleon axial charge
has been studied within  HB$\chi$PT in the SSE formulation~\cite{Hemmert:2003cb}.  
We use the one-loop result including explicitly the $\Delta$
degrees of freedom in order to extrapolate our lattice results to the
physical point. We make a three parameter fit to the form 

\beq
g_A(m_\pi^2) &=& g_A^0-\frac{(g_A^0)^3 m_\pi^2}{16 \pi^2 f_\pi^2}+
4\Biggl\{C_{SSE}(\lambda) \nonumber \\
&{}&+\frac{c_A^2}{4\pi^2 f_\pi^2}\left[\frac{155}{972}g_1 -\frac{17}{36}g_A^0\right] + \gamma\ln\frac{m_\pi}{\lambda}\Biggr\} m_\pi^2
\nonumber \\
& {}& +\frac{4c_A^2g_a^0}{27\pi f_\pi^2 \Delta_0}m_\pi^2+\frac{8}{27\pi^2 f_\pi^2}c_A^2g_A^0 m_\pi^2 R(m_\pi) \nonumber \\
&{}& + \frac{c_A^2\Delta_0^2}{81\pi^2f_\pi^2}(25 g_1-57g_A^0) \Biggl\{\ln\frac{2\Delta_0}{m_\pi}-R(m_\pi)\Biggr\} \nonumber \\
&{}&+{\cal O}(\epsilon^4)\, ,
\eeq
with 
\beq 
\gamma &=& \frac{1}{16\pi^2 f_\pi^2}\left[\frac{50}{81}c_A^2g_1-\frac{1}{2}g_A^0
-\frac{2}{9}c_A^2 g_A^0 -(g_A^0)^3\right] \nonumber\\
R(m_\pi)&=&\sqrt{1-\frac{m_\pi}{\Delta_0}}\left[\frac{\Delta_0}{m_\pi}+\sqrt{\frac{\Delta_0^2}{m_\pi^2}-1}\right].
\eeq

The three parameters to
fit are  $g_A^0$, the value of the axial charge at the chiral point, 
the $\Delta$ axial coupling constant $g_1$ and a counter-term  $C_{SSE}(\lambda)$. 
We again take the nucleon to $\Delta$ axial coupling constant $c_A=1.5$, 
the mass splitting between the $\Delta$ and the nucleon at the chiral
limit, $\Delta_0=0.2711$ and $\lambda= 1$~GeV.
Fitting  the volume corrected  continuum results we find
$g_A{=}1.12(8)$, $g_1{=}2.37(1.52)$ and
$C_{SSE}{=}-1.01(2.01)$. The parameters $g_1$ and $C_{SSE}$ are highly
correlated explaining the resulting large error band. Fitting the lattice
without any volume correction and without extrapolating to the continuum
limit we obtained $g_A{=}1.08(8)$, $g_1{=}2.02(1.21)$ and
$C_{SSE}{=}-0.63(1.57)$ which are consistent with the continuum volume
corrected results. This shows that both cut-off and volume artifacts are
small as compared to the uncertainty due to the chiral extrapolation.

\subsection{Axial form factors}
In this section we discuss the results obtained for the axial form factors $G_A(Q^2)$ and $G_p(Q^2)$.

\begin{figure}
\includegraphics[width=\linewidth]{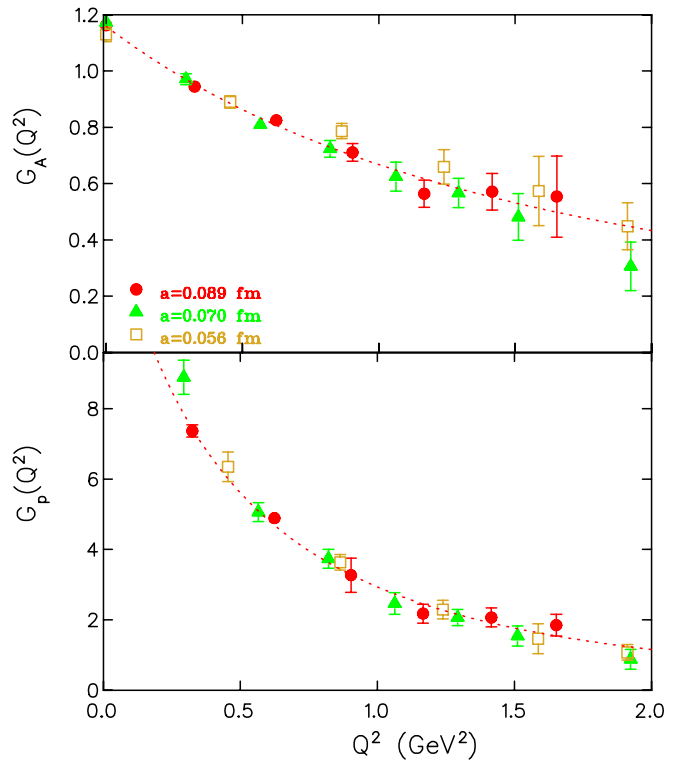}
\caption{The nucleon axial form factors $G_A(Q^2)$ and $G_p(Q^2)$ at $m_\pi\sim 470$~MeV
at $\beta=3.9$ (filled red circles), $4.05$ (filled green triangles) and $4.2$ (yellow squares) versus $Q^2$. The  line is the result of a dipole fit (to the form given in Eq.~(\ref{Gp_fit}))
$G_A(Q^2)$ ($G_p(Q^2)$)  data
 on the coarse lattice. }
\label{fig:GAGp a}
\end{figure}

To assess cut-off effects we compare in Fig.~\ref{fig:GAGp a} results for
 $G_A(Q^2)$ and $G_P(Q^2)$ versus 
$Q^2$
for three different lattice spacings at a similar pion mass of about $470$~MeV.
As can be seen, results at these three lattice spacings are consistent indicating
that cut-off effects are negligible for these lattice spacings. 
We perform a dipole fit to $G_A(Q^2)$ using
\be
G_A(Q^2){=}\frac{g_A}{(1{+}Q^2/m_A^2)^2}\,,
\label{GA_fit}
\ee
with a momentum upper  range of $Q^2\sim 1.5$GeV$^2$.
The  axial mass $m_A$ of the fits is larger than in
experimental value of $m_A^{\rm exp}=1.1$~GeV  extracted from the best dipole
fit to the electroproduction data. This is  
evident from the smaller slope shown by the lattice
data  both for twisted mass fermions and
domain wall fermions.
Assuming the partially conserved axial current relation and pion pole dominance we can relate the form factor $G_p(Q^2)$ to $G_A(Q^2)$:
\be
G_p(Q^2){=}G_A(Q^2) \frac{G_p(0)}{Q^2+m_p^2}\,.
\label{Gp_fit}
\ee

\begin{figure}
\includegraphics[width=\linewidth]{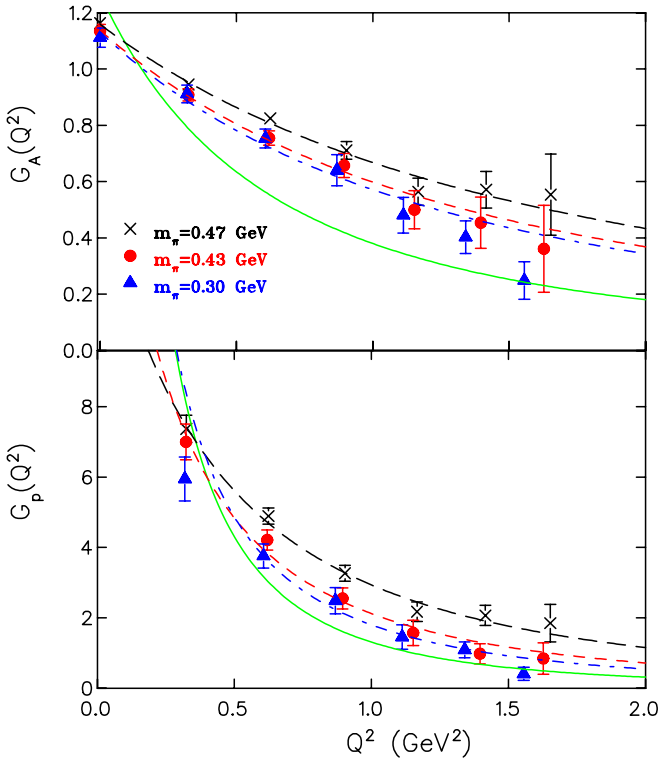}            
\caption{The nucleon axial form factors $G_A(Q^2)$ and $G_p(Q^2)$ 
at $\beta=3.9$ for $m_\pi = 468 $~MeV (crosses), $m_\pi = 432 $~MeV (filled red circles) and $m_\pi = 303 $~MeV (filled blue triangles) versus $Q^2$. The dashed lines are the result of a dipole fit for 
$G_A(Q^2)$  and to the form given in Eq.~(\ref{Gp_fit})  for $G_p(Q^2)$ 
 on the coarse lattice.}
\label{fig:GAGp mpi}
\end{figure}

The  dependence of these form factors
on the pion mass is seen in Fig.~\ref{fig:GAGp mpi}, where we
show both $G_A$ and $G_p$ computed at several values of the pion mass
spanning a range from about 470~MeV to 300~MeV at $\beta=3.9$.
 We show fits to the lattice data using a dipole form as given in Eq.~(\ref{GA_fit}) for $G_A(Q^2)$ and to the form given in Eq.~(\ref{Gp_fit}) for $G_p(Q^2)$, which described the data rather well.
 Although  the mass dependence is weak and the general trend is to approach
experiment, lattice data  show a weaker $Q^2$ dependence 
as compared to experiment.  As already pointed out, the best
dipole fit to the electroproduction data yields an axial mass $m_A^{\rm exp}= 1.1$~GeV~\cite{Bernard:2001rs}, and it is shown by the solid line. The experimental line in the
case of $G_P(Q^2)$ shown in Fig.~\ref{fig:GAGp mpi} is obtained using  Eq.~(\ref{Gp_fit}) and  pion 
mass of $m_p=135$~MeV. 

\begin{figure}
\includegraphics[width=\linewidth]{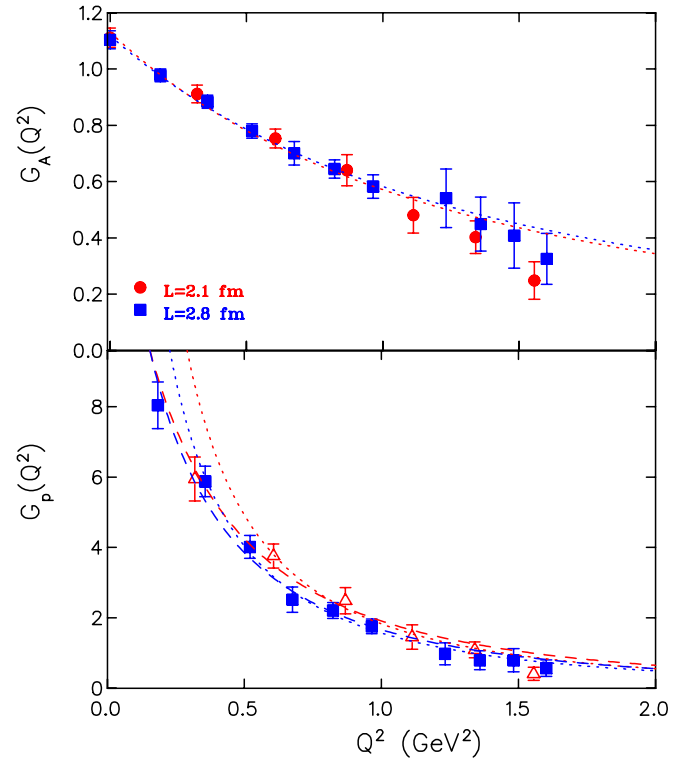}
\caption{The nucleon axial form factors $G_A$ and $G_p$ at $m_\pi\sim 300$~MeV
for a lattice of size $24^3\times 48$ (filled red circles) and 
$32^3\times 64$ (filled blue squares). For $G_A(Q^2)$ (upper graph) the dotted  lines are the best dipole fits to the lattice data. For
$G_p(Q^2)$ (lower graph)
 the dotted lines are  fits of lattice
results to the form $\frac{CG_A(Q^2)}{(Q^2+m_p^2)}$ discarding 
the point at the lowest value of $Q^2$. The dashed
lines are fits using all data points.}
\label{fig:GAGp L}
\end{figure}
 \begin{figure}
\includegraphics[width=\linewidth]{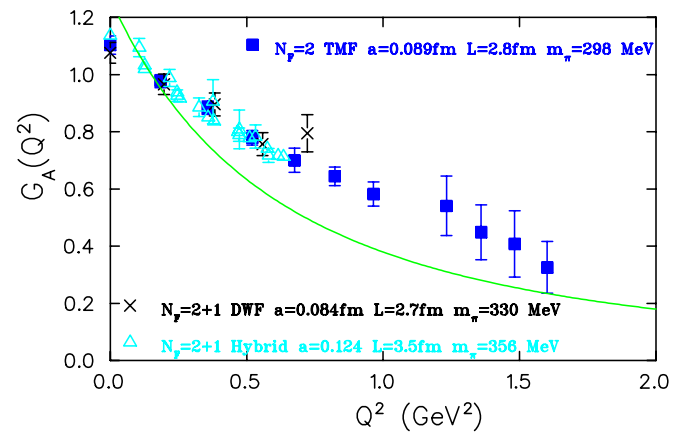}
\caption{ Axial form factor $G_A(Q^2)$ as a function of $Q^2$. $N_F=2$ TMF
results at $m_\pi=298$~MeV  are shown with filled squares, $N_F=2+1$ DWF 
with the crosses at $m_\pi=330$~Mev and $N_F=2+1$ using a mixed action
of DWF and staggered sea quarks at $m_\pi=356$~MeV with triangles. The solid line is the dipole parametrization of experimental data. }
\label{fig:GA}
\end{figure}
\begin{figure}
\includegraphics[width=\linewidth]{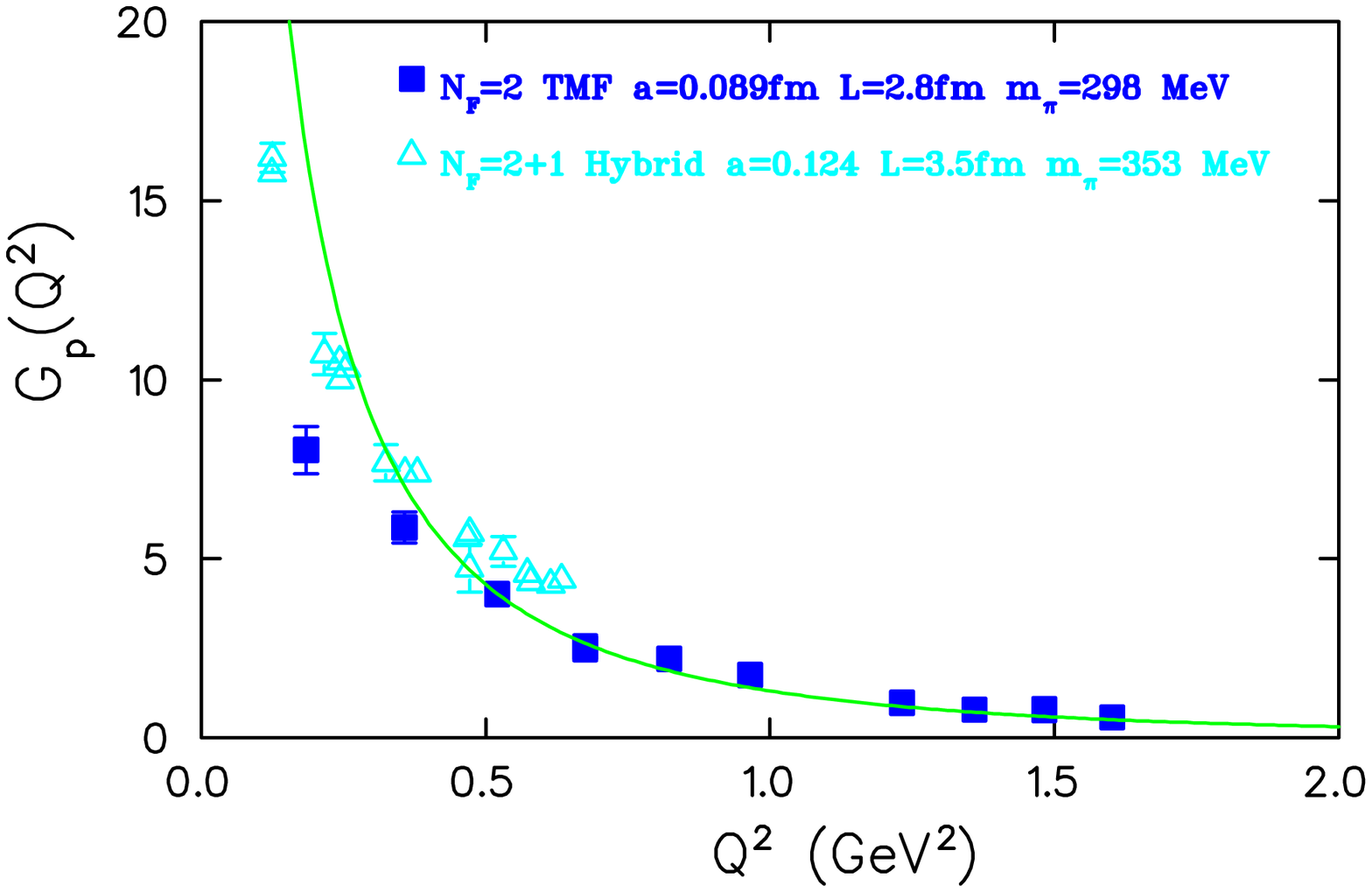}
\caption{Induced pseudo-scalar  form factor $G_p(Q^2)$ as a function of $Q^2$. $N_F=2$ TMF
results at $m_\pi=298$~MeV  are shown with filled squares and $N_F=2+1$ using a mixed action
of DWF and staggered sea quarks at $m_\pi=356$~MeV are shown  with triangles.
 The solid line is obtained using the parametrization of experimental electroproduction results  for $G_A$
and pion-pole dominance.}
\label{fig:Gp}
   \end{figure}

In Fig.~\ref{fig:GAGp L} we check for finite volume effects by comparing results obtained at $\beta=3.9$ 
on a lattice of spatial length $L=2.8$~fm and $L=2.1$~fm at  $m_\pi \sim 300$~MeV. As can be seen
volume effects are negligible for $G_A$. 
In the case of $G_p$ we have a strong dependence on $Q^2$ as $Q^2\rightarrow 0$ 
because of the pion pole dependence expected for this form factor. 
Therefore the fits are strongly dependent on the
lowest values of $Q^2$ that are available. E.g. discarding the point at the lowest momentum yield the dotted lines which are steeper as compared to including it.
Although there is an overall consistency between 
the two data sets at $\beta=3.9$ deviations are seen in the fits when
the same momentum range is used. The fit using the whole
range of data obtained on the smaller lattice, shown by the dashed red line
exhibit
a weaker dependence as compared to the fit using the results on the
larger lattice but discarding the point at the lowest momentum, shown by the dotted blue line. 
In  Figs.~\ref{fig:GA} and~\ref{fig:Gp} we compare our results with the results 
by LHPC which were obtained in a mixed action approach that uses 
DWF on staggered sea quarks~\cite{Bratt:2010jn}  on a lattice
with $L=3.5$~fm. The results are in   agreement in the case of $G_A(Q^2)$,
while in the 
case of $G_p(Q^2)$ there are larger discrepancies. Given the mass dependence of $G_p(Q^2)$ shown in Fig.~\ref{fig:GAGp mpi}
a difference in the pion mass of  50~MeV cannot fully account for this.
Such discrepancies may indicate  that volume effects are not negligible 
on form factors such as 
$G_p(Q^2)$ which are strongly affected  by the pion-pole.

\begin{table}[h]
\begin{center}
\begin{tabular}{c|c|c|c|c|c}
\hline\hline
$a\mu$ & $m_\pi$ (GeV)  & $g_A$ & $m_A$ (GeV)& $G_p(0)$ & $m_{p}$ (GeV)  \\\hline 
\multicolumn{6}{c}{$\beta=3.9$}\\\hline 
0.0100 & 0.4675 &    1.163(14) & 1.776(48)   &  6.99(74) &  0.738(77)  \\
0.0085 & 0.4319 &    1.140(11) & 1.634(32)   & 4.40(62)  & 0.458(106)  \\
0.0064 & 0.3770 &    1.081(31) & 2.021(17)   & 3.33(40)  & 0.254(113)  \\
0.004  & 0.3032 &    1.135(34) & 1.572(82)   & 4.30(76)  & 0.512(133)  \\
0.004  & 0.2978 &    1.160(37) & 1.513(70)   & 4.28(63)  & 0.459(119)  \\
0.003  & 0.2600 &    1.166(28) & 1.445(51)   & 2.80(25)  & 0.255(95)    \\\hline
\multicolumn{6}{c}{$\beta=4.05$}\\\hline 
0.008 & 0.4653   &   1.174(24)  & 1.696(24) &   5.88(43) &  0.595(53) \\
0.006 & 0.4035   &   1.174(32)  & 1.725(65) &   4.90(84) &  0.514(124)\\
0.003 & 0.2925   &   1.180(67)  & 1.392(44) &   4.13(85) &  0.458(153)\\\hline
\multicolumn{6}{c}{$\beta=4.2$}\\\hline 
0.0065& 0.4698 &      1.120(21)  & 2.030(93) &    5.72(40) &   0.599(60) \\
0.002 & 0.2622 &      1.158(22)  & 1.575(42) &    3.56(18) &   0.325(39) \\\hline
\end{tabular}
\caption{Results on the axial nucleon charge and axial mass extracted by fitting
$G_A(Q^2)$ to a dipole form. The two last column give the $G_p(0)$ and  the mass $m_p$ by fitting $G_p(Q^2)$ to the form given in Eq.~(\ref{Gp_fit}).}
\label{Table:GA Gp fits}
\end{center}
\end{table}

In the case of $G_p(Q^2)$ one can extract the fit parameters $G_p(0)$ and $m_p$
by  either fitting  the ratio of $G_p/G_A$ 
 or
 use the fitted values for
$G_A(q^2)$ and fit to the form of Eq.~(\ref{Gp_fit}). 
The values extracted by performing these fits are compatible within
error bars. 
 Our lattice data on
$G_p(q^2)$ are flatter than pion-pole dominance predicts requiring a
larger pole mass $m_p$ than the pion mass measured on the lattice. 
In Table~\ref{Table:GA Gp fits} we tabulate the resulting fitting
parameters for all $\beta$ and $\mu$ values. The parameters $G_p(0)$
and $m_p$ have been extracted from
fits to the form given in Eq.~(\ref{Gp_fit}).
The full set of our lattice results on $G_A(q^2)$ and $G_p(Q^2)$ is given in the Tables~VI-VIII
of the Appendix.
\section{Conclusions}
Using $N_F=2$ twisted mass fermions we obtain accurate results on the
axial $G_A(Q^2)$, and $G_p(Q^2)$ form factors as a function of $Q^2$
for pion masses in the range of about  260-470 MeV. The general feature
is a flatter dependence on $Q^2$ than experiment. This is a feature
also seen in the electromagnetic nucleon form factors. Finite volume
effects are found to be small on $G_A$. Our results
are in agreement with recent results obtained using  other lattice
fermions such as dynamical
$N_F=2+1$ domain wall fermions. 
Having results at three lattice spacings
enables us to take the continuum limit.
We find that cut-off effects are small for the values of the lattice spacings used in this work. Performing a chiral extrapolation of our continuum
results for the nucleon axial charge, we find
at the physical point  the value $g_A=1.12(8)$. This is one
standard deviation lower than the physical value. The large error associated
with our determination of $g_A$ is
mostly due to the chiral extrapolation. Therefore it is crucial to 
perform an analysis with a pion mass closer to its physical value.

\section*{Acknowledgments}
We would like to thank all members of ETMC for a
very constructive and enjoyable collaboration and for the many fruitful 
discussions that took place during the development of this work.

Numerical calculations have  used HPC resources from GENCI (IDRIS and CINES) Grant 2009-052271  and
CC-IN2P3 as well as  from the
John von Neumann-Institute for Computing on the JUMP and  Jugene systems at the 
research center
in J\"ulich. We thank the staff members for their kind and sustained support.
M.Papinutto acknowledges financial support by the Marie Curie European
Reintegration Grant of the 7th European Community Framework
Programme under contract number PERG05-GA-2009-249309. 
This work is supported in part by  the DFG
Sonder\-for\-schungs\-be\-reich/ Trans\-regio SFB/TR9 and by funding received from the
 Cyprus Research Promotion Foundation under contracts EPYAN/0506/08,
KY-$\Gamma$/0907/11/ and TECHNOLOGY/$\Theta$E$\Pi$I$\Sigma$/0308(BE)/17. 

\bibliography{axial_ref}

\begin{thebibliography}{34}
\expandafter\ifx\csname natexlab\endcsname\relax\def\natexlab#1{#1}\fi
\expandafter\ifx\csname bibnamefont\endcsname\relax
  \def\bibnamefont#1{#1}\fi
\expandafter\ifx\csname bibfnamefont\endcsname\relax
  \def\bibfnamefont#1{#1}\fi
\expandafter\ifx\csname citenamefont\endcsname\relax
  \def\citenamefont#1{#1}\fi
\expandafter\ifx\csname url\endcsname\relax
  \def\url#1{\texttt{#1}}\fi
\expandafter\ifx\csname urlprefix\endcsname\relax\def\urlprefix{URL }\fi
\providecommand{\bibinfo}[2]{#2}
\providecommand{\eprint}[2][]{\url{#2}}

\bibitem[{\citenamefont{Ahrens et~al.}(1988)}]{Ahrens:1988rr}
\bibinfo{author}{\bibfnamefont{L.~A.} \bibnamefont{Ahrens}}
  \bibnamefont{et~al.}, \bibinfo{journal}{Phys. Lett.}
  \textbf{\bibinfo{volume}{B202}}, \bibinfo{pages}{284} (\bibinfo{year}{1988}).

\bibitem[{\citenamefont{Bernard et~al.}(1992)\citenamefont{Bernard, Kaiser, and
  Meissner}}]{Bernard:1992ys}
\bibinfo{author}{\bibfnamefont{V.}~\bibnamefont{Bernard}},
  \bibinfo{author}{\bibfnamefont{N.}~\bibnamefont{Kaiser}}, \bibnamefont{and}
  \bibinfo{author}{\bibfnamefont{U.~G.} \bibnamefont{Meissner}},
  \bibinfo{journal}{Phys. Rev. Lett.} \textbf{\bibinfo{volume}{69}},
  \bibinfo{pages}{1877} (\bibinfo{year}{1992}).

\bibitem[{\citenamefont{Choi et~al.}(1993)}]{Choi:1993vt}
\bibinfo{author}{\bibfnamefont{S.}~\bibnamefont{Choi}} \bibnamefont{et~al.},
  \bibinfo{journal}{Phys. Rev. Lett.} \textbf{\bibinfo{volume}{71}},
  \bibinfo{pages}{3927} (\bibinfo{year}{1993}).

\bibitem[{\citenamefont{Gorringe and Fearing}(2004)}]{Gorringe:2002xx}
\bibinfo{author}{\bibfnamefont{T.}~\bibnamefont{Gorringe}} \bibnamefont{and}
  \bibinfo{author}{\bibfnamefont{H.~W.} \bibnamefont{Fearing}},
  \bibinfo{journal}{Rev. Mod. Phys.} \textbf{\bibinfo{volume}{76}},
  \bibinfo{pages}{31} (\bibinfo{year}{2004}), \eprint{nucl-th/0206039}.

\bibitem[{\citenamefont{Bernard et~al.}(2002)\citenamefont{Bernard,
  Elouadrhiri, and Meissner}}]{Bernard:2001rs}
\bibinfo{author}{\bibfnamefont{V.}~\bibnamefont{Bernard}},
  \bibinfo{author}{\bibfnamefont{L.}~\bibnamefont{Elouadrhiri}},
  \bibnamefont{and} \bibinfo{author}{\bibfnamefont{U.~G.}
  \bibnamefont{Meissner}}, \bibinfo{journal}{J. Phys.}
  \textbf{\bibinfo{volume}{G28}}, \bibinfo{pages}{R1} (\bibinfo{year}{2002}),
  \eprint{hep-ph/0107088}.

\bibitem[{\citenamefont{Schindler et~al.}(2007)\citenamefont{Schindler, Fuchs,
  Gegelia, and Scherer}}]{Schindler:2006it}
\bibinfo{author}{\bibfnamefont{M.~R.} \bibnamefont{Schindler}},
  \bibinfo{author}{\bibfnamefont{T.}~\bibnamefont{Fuchs}},
  \bibinfo{author}{\bibfnamefont{J.}~\bibnamefont{Gegelia}}, \bibnamefont{and}
  \bibinfo{author}{\bibfnamefont{S.}~\bibnamefont{Scherer}},
  \bibinfo{journal}{Phys. Rev.} \textbf{\bibinfo{volume}{C75}},
  \bibinfo{pages}{025202} (\bibinfo{year}{2007}), \eprint{nucl-th/0611083}.

\bibitem[{\citenamefont{Shindler}(2008)}]{Shindler:2007vp}
\bibinfo{author}{\bibfnamefont{A.}~\bibnamefont{Shindler}},
  \bibinfo{journal}{Phys. Rept.} \textbf{\bibinfo{volume}{461}},
  \bibinfo{pages}{37} (\bibinfo{year}{2008}), \eprint{0707.4093}.

\bibitem[{\citenamefont{Frezzotti et~al.}(2001)\citenamefont{Frezzotti, Grassi,
  Sint, and Weisz}}]{Frezzotti:2000nk}
\bibinfo{author}{\bibfnamefont{R.}~\bibnamefont{Frezzotti}},
  \bibinfo{author}{\bibfnamefont{P.~A.} \bibnamefont{Grassi}},
  \bibinfo{author}{\bibfnamefont{S.}~\bibnamefont{Sint}}, \bibnamefont{and}
  \bibinfo{author}{\bibfnamefont{P.}~\bibnamefont{Weisz}}
  (\bibinfo{collaboration}{Alpha}), \bibinfo{journal}{JHEP}
  \textbf{\bibinfo{volume}{0108}}, \bibinfo{pages}{058} (\bibinfo{year}{2001}),
  \eprint{hep-lat/0101001}.

\bibitem[{\citenamefont{Weisz}(1983)}]{Weisz:1982zw}
\bibinfo{author}{\bibfnamefont{P.}~\bibnamefont{Weisz}},
  \bibinfo{journal}{Nucl. Phys.} \textbf{\bibinfo{volume}{B212}},
  \bibinfo{pages}{1} (\bibinfo{year}{1983}).

\bibitem[{\citenamefont{Alexandrou}(2009)}]{Alexandrou:2009xk}
\bibinfo{author}{\bibfnamefont{C.}~\bibnamefont{Alexandrou}}
  (\bibinfo{year}{2009}), \eprint{0906.4137}.

\bibitem[{\citenamefont{Alexandrou
  et~al.}(2009{\natexlab{a}})}]{Alexandrou:2009qu}
\bibinfo{author}{\bibfnamefont{C.}~\bibnamefont{Alexandrou}}
  \bibnamefont{et~al.} (\bibinfo{collaboration}{ETM}), \bibinfo{journal}{Phys.
  Rev.} \textbf{\bibinfo{volume}{D80}}, \bibinfo{pages}{114503}
  (\bibinfo{year}{2009}{\natexlab{a}}), \eprint{0910.2419}.

\bibitem[{\citenamefont{Drach et~al.}(2008)}]{Drach:2009dh}
\bibinfo{author}{\bibfnamefont{V.}~\bibnamefont{Drach}} \bibnamefont{et~al.},
  \bibinfo{journal}{PoS} \textbf{\bibinfo{volume}{LATTICE2008}},
  \bibinfo{pages}{123} (\bibinfo{year}{2008}), \eprint{0905.2894}.

\bibitem[{\citenamefont{Alexandrou
  et~al.}(2008{\natexlab{a}})}]{Alexandrou:2008tn}
\bibinfo{author}{\bibfnamefont{C.}~\bibnamefont{Alexandrou}}
  \bibnamefont{et~al.} (\bibinfo{collaboration}{European Twisted Mass}),
  \bibinfo{journal}{Phys. Rev.} \textbf{\bibinfo{volume}{D78}},
  \bibinfo{pages}{014509} (\bibinfo{year}{2008}{\natexlab{a}}),
  \eprint{0803.3190}.

\bibitem[{\citenamefont{Alexandrou et~al.}(2007)}]{Alexandrou:2007qq}
\bibinfo{author}{\bibfnamefont{C.}~\bibnamefont{Alexandrou}}
  \bibnamefont{et~al.} (\bibinfo{collaboration}{ETM Collaboration}),
  \bibinfo{journal}{PoS} \textbf{\bibinfo{volume}{LAT2007}},
  \bibinfo{pages}{087} (\bibinfo{year}{2007}), \eprint{arXiv:0710.1173
  [hep-lat]}.

\bibitem[{\citenamefont{Drach et~al.}(2010)}]{Drach:2010}
\bibinfo{author}{\bibfnamefont{V.}~\bibnamefont{Drach}} \bibnamefont{et~al.},
  \bibinfo{journal}{PoS} \textbf{\bibinfo{volume}{Lattice 2010}},
  \bibinfo{pages}{123} (\bibinfo{year}{2010}).

\bibitem[{\citenamefont{Alexandrou}(2010)}]{Alexandrou:2010}
\bibinfo{author}{\bibfnamefont{C.}~\bibnamefont{Alexandrou}}
  (\bibinfo{collaboration}{ETM Collaboration}), \bibinfo{journal}{PoS}
  \textbf{\bibinfo{volume}{Lattice 2010}}, \bibinfo{pages}{001}
  (\bibinfo{year}{2010}).

\bibitem[{\citenamefont{Alexandrou
  et~al.}(2009{\natexlab{b}})}]{Alexandrou:2009ng}
\bibinfo{author}{\bibfnamefont{C.}~\bibnamefont{Alexandrou}}
  \bibnamefont{et~al.}, \bibinfo{journal}{PoS}
  \textbf{\bibinfo{volume}{LAT2009}}, \bibinfo{pages}{145}
  (\bibinfo{year}{2009}{\natexlab{b}}), \eprint{0910.3309}.

\bibitem[{\citenamefont{Alexandrou
  et~al.}(2008{\natexlab{b}})}]{Alexandrou:2008rp}
\bibinfo{author}{\bibfnamefont{C.}~\bibnamefont{Alexandrou}}
  \bibnamefont{et~al.}, \bibinfo{journal}{PoS LAT2008}
  \textbf{\bibinfo{volume}{B414}}, \bibinfo{pages}{145}
  (\bibinfo{year}{2008}{\natexlab{b}}), \eprint{hep-lat/9211042}.

\bibitem[{\citenamefont{Dimopoulos et~al.}(2007)\citenamefont{Dimopoulos,
  Frezzotti, Herdoiza, Urbach, and Wenger}}]{Dimopoulos:2007qy}
\bibinfo{author}{\bibfnamefont{P.}~\bibnamefont{Dimopoulos}},
  \bibinfo{author}{\bibfnamefont{R.}~\bibnamefont{Frezzotti}},
  \bibinfo{author}{\bibfnamefont{G.}~\bibnamefont{Herdoiza}},
  \bibinfo{author}{\bibfnamefont{C.}~\bibnamefont{Urbach}}, \bibnamefont{and}
  \bibinfo{author}{\bibfnamefont{U.}~\bibnamefont{Wenger}}
  (\bibinfo{collaboration}{ETM Collaboration}), \bibinfo{journal}{PoS}
  \textbf{\bibinfo{volume}{LAT2007}} (\bibinfo{year}{2007}),
  \eprint{arXiv:0710.2498 [hep-lat]}.

\bibitem[{\citenamefont{Constantinou
  et~al.}(2010{\natexlab{a}})}]{Constantinou:2010gr}
\bibinfo{author}{\bibfnamefont{M.}~\bibnamefont{Constantinou}}
  \bibnamefont{et~al.} (\bibinfo{year}{2010}{\natexlab{a}}),
  \eprint{1004.1115}.

\bibitem[{\citenamefont{Alexandrou et~al.}(2010)\citenamefont{Alexandrou,
  Constantinou, Korzec, Panagopoulos, and Stylianou}}]{Alexandrou:2010me}
\bibinfo{author}{\bibfnamefont{C.}~\bibnamefont{Alexandrou}},
  \bibinfo{author}{\bibfnamefont{M.}~\bibnamefont{Constantinou}},
  \bibinfo{author}{\bibfnamefont{T.}~\bibnamefont{Korzec}},
  \bibinfo{author}{\bibfnamefont{H.}~\bibnamefont{Panagopoulos}},
  \bibnamefont{and} \bibinfo{author}{\bibfnamefont{F.}~\bibnamefont{Stylianou}}
  (\bibinfo{year}{2010}), \eprint{1006.1920}.

\bibitem[{\citenamefont{Constantinou
  et~al.}(2010{\natexlab{b}})\citenamefont{Constantinou, Panagopoulos, and
  Stylianou}}]{Constantinou:2010dn}
\bibinfo{author}{\bibfnamefont{M.}~\bibnamefont{Constantinou}},
  \bibinfo{author}{\bibfnamefont{H.}~\bibnamefont{Panagopoulos}},
  \bibnamefont{and} \bibinfo{author}{\bibfnamefont{F.}~\bibnamefont{Stylianou}}
  (\bibinfo{year}{2010}{\natexlab{b}}), \eprint{1001.1498}.

\bibitem[{\citenamefont{Gockeler et~al.}(1999)}]{Gockeler:1998ye}
\bibinfo{author}{\bibfnamefont{M.}~\bibnamefont{Gockeler}}
  \bibnamefont{et~al.}, \bibinfo{journal}{Nucl. Phys.}
  \textbf{\bibinfo{volume}{B544}}, \bibinfo{pages}{699} (\bibinfo{year}{1999}),
  \eprint{hep-lat/9807044}.

\bibitem[{\citenamefont{Alexandrou et~al.}()\citenamefont{Alexandrou,
  Constantinou, Korzec, Panagopoulos, and Stylianou}}]{Martha}
\bibinfo{author}{\bibfnamefont{C.}~\bibnamefont{Alexandrou}},
  \bibinfo{author}{\bibfnamefont{M.}~\bibnamefont{Constantinou}},
  \bibinfo{author}{\bibfnamefont{T.}~\bibnamefont{Korzec}},
  \bibinfo{author}{\bibfnamefont{H.}~\bibnamefont{Panagopoulos}},
  \bibnamefont{and}
  \bibinfo{author}{\bibfnamefont{F.}~\bibnamefont{Stylianou}},
  \bibinfo{note}{poS Latt2010, Sardinia, Itlay, 14-19 Jun 2010; in
  preparation}.

\bibitem[{\citenamefont{Alexandrou
  et~al.}(2009{\natexlab{c}})}]{Alexandrou:GPDs}
\bibinfo{author}{\bibfnamefont{C.}~\bibnamefont{Alexandrou}}
  \bibnamefont{et~al.} (\bibinfo{collaboration}{ETM Collaboration}),
  \bibinfo{journal}{PoS} \textbf{\bibinfo{volume}{LAT2009}},
  \bibinfo{pages}{136} (\bibinfo{year}{2009}{\natexlab{c}}).

\bibitem[{\citenamefont{Alexandrou et~al.}(1994)\citenamefont{Alexandrou,
  Gusken, Jegerlehner, Schilling, and Sommer}}]{Alexandrou:1992ti}
\bibinfo{author}{\bibfnamefont{C.}~\bibnamefont{Alexandrou}},
  \bibinfo{author}{\bibfnamefont{S.}~\bibnamefont{Gusken}},
  \bibinfo{author}{\bibfnamefont{F.}~\bibnamefont{Jegerlehner}},
  \bibinfo{author}{\bibfnamefont{K.}~\bibnamefont{Schilling}},
  \bibnamefont{and} \bibinfo{author}{\bibfnamefont{R.}~\bibnamefont{Sommer}},
  \bibinfo{journal}{Nucl. Phys.} \textbf{\bibinfo{volume}{B414}},
  \bibinfo{pages}{815} (\bibinfo{year}{1994}), \eprint{hep-lat/9211042}.

\bibitem[{\citenamefont{Gusken}(1990)}]{Gusken:1989}
\bibinfo{author}{\bibfnamefont{S.}~\bibnamefont{Gusken}},
  \bibinfo{journal}{Nucl. Phys. Proc. Suppl.} \textbf{\bibinfo{volume}{17}},
  \bibinfo{pages}{361} (\bibinfo{year}{1990}).

\bibitem[{\citenamefont{Urbach}(2007)}]{Urbach:2007}
\bibinfo{author}{\bibfnamefont{C.}~\bibnamefont{Urbach}},
  \bibinfo{journal}{PoS} \textbf{\bibinfo{volume}{LAT2007}},
  \bibinfo{pages}{022} (\bibinfo{year}{2007}).

\bibitem[{\citenamefont{Ali~Khan et~al.}(2004)}]{AliKhan:2003cu}
\bibinfo{author}{\bibfnamefont{A.}~\bibnamefont{Ali~Khan}} \bibnamefont{et~al.}
  (\bibinfo{collaboration}{QCDSF-UKQCD}), \bibinfo{journal}{Nucl. Phys.}
  \textbf{\bibinfo{volume}{B689}}, \bibinfo{pages}{175} (\bibinfo{year}{2004}),
  \eprint{hep-lat/0312030}.

\bibitem[{\citenamefont{Yamazaki et~al.}(2009)}]{Yamazaki:2009zq}
\bibinfo{author}{\bibfnamefont{T.}~\bibnamefont{Yamazaki}}
  \bibnamefont{et~al.}, \bibinfo{journal}{Phys. Rev.}
  \textbf{\bibinfo{volume}{D79}}, \bibinfo{pages}{114505}
  (\bibinfo{year}{2009}), \eprint{0904.2039}.

\bibitem[{\citenamefont{Bratt et~al.}(2010)}]{Bratt:2010jn}
\bibinfo{author}{\bibfnamefont{J.~D.} \bibnamefont{Bratt}} \bibnamefont{et~al.}
  (\bibinfo{collaboration}{LHPC}) (\bibinfo{year}{2010}), \eprint{1001.3620}.

\bibitem[{\citenamefont{Hemmert et~al.}(1998)\citenamefont{Hemmert, Holstein,
  and Kambor}}]{Hemmert:1997ye}
\bibinfo{author}{\bibfnamefont{T.~R.} \bibnamefont{Hemmert}},
  \bibinfo{author}{\bibfnamefont{B.~R.} \bibnamefont{Holstein}},
  \bibnamefont{and} \bibinfo{author}{\bibfnamefont{J.}~\bibnamefont{Kambor}},
  \bibinfo{journal}{J. Phys.} \textbf{\bibinfo{volume}{G24}},
  \bibinfo{pages}{1831} (\bibinfo{year}{1998}), \eprint{hep-ph/9712496}.

\bibitem[{\citenamefont{Khan et~al.}(2006)}]{Khan:2006de}
\bibinfo{author}{\bibfnamefont{A.~A.} \bibnamefont{Khan}} \bibnamefont{et~al.},
  \bibinfo{journal}{Phys. Rev.} \textbf{\bibinfo{volume}{D74}},
  \bibinfo{pages}{094508} (\bibinfo{year}{2006}), \eprint{hep-lat/0603028}.

\bibitem[{\citenamefont{Hemmert et~al.}(2003)\citenamefont{Hemmert, Procura,
  and Weise}}]{Hemmert:2003cb}
\bibinfo{author}{\bibfnamefont{T.~R.} \bibnamefont{Hemmert}},
  \bibinfo{author}{\bibfnamefont{M.}~\bibnamefont{Procura}}, \bibnamefont{and}
  \bibinfo{author}{\bibfnamefont{W.}~\bibnamefont{Weise}},
  \bibinfo{journal}{Phys. Rev.} \textbf{\bibinfo{volume}{D68}},
  \bibinfo{pages}{075009} (\bibinfo{year}{2003}), \eprint{hep-lat/0303002}.

\end{thebibliography}
\begin{widetext}
\section{Appendix}
\begin{table}[h]
\begin{center}
\begin{tabular}{c|c|c|c|}
\hline\hline
 $m_\pi$ (GeV)  & $(aQ)^2$  & $G_A$ & $G_p$    \\
 (no. confs)   &   &  &     \\\hline 
\multicolumn{4}{c}{$\beta=3.9$, $24^3\times 48$ }\\\hline 
            &0.0      &  1.163(18) &             \\
            & 0.066   &  0.945(15) &  7.368(390) \\
   0.4675    & 0.126   &  0.825(19) &  4.889(230) \\
  (477)     & 0.182   &  0.711(31) &  3.267(227) \\
            & 0.235(1)&  0.564(48) &  2.175(278) \\
            & 0.286(1)&  0.571(65) &  2.066(287) \\
            & 0.334(1)&  0.554(14) &  1.847(536) \\
 \hline
            &0.0      &   1.134(25)  &             \\
            &0.065    &   0.908(22)  &  6.999(510)  \\
0.4319       &0.125    &   0.755(26)  &  4.211(287) \\
(365)       &0.181(1) &   0.657(43)  &  2.551(298)  \\
            &0.233(1) &   0.499(68)  &  1.571(361) \\
            &0.282(1) &   0.454(91)  &  0.977(288) \\
            &0.328(2) &   0.361(154) &  0.845(445) \\
 \hline
        &  0.0    & 1. 140(27) &              \\
        &  0.065  &  0.931(24) &  7.504(614)   \\
  0.3770 & 0.125   &  0.788(25)&  4.145(325)   \\
  (553) & 0.180(1)& 0.737(74)  &  3.092(453)   \\
        & 0.231(1)& 0.648(211) &  2.352(916)   \\
        & 0.280(1)& 0.631(202) &  1.860(687)  \\
        & 0.326(2)& 0.329(292) &  0.844(872)  \\
\hline
       &0.0        & 1.111(34)  &               \\
       &0.064      & 0.911(31)  &   5.947(626)  \\
 0.3032 & 0.122     & 0.753(33) &   3.757(341)  \\
 (943) &0.175(1)   & 0.640(55)  &   2.486(370)  \\
       &0.224(1)   & 0.480(63)  &   1.454(351)  \\
       & 0.270(2)  & 0.402(58)  &   1.090(223)  \\
       &0.314(2)   & 0.248(67)  &   0.404(186)  \\
\hline
\multicolumn{4}{c}{$\beta=3.9$, $32^3\times 64$ }\\\hline 
      & 0.00      &  1.103(32)  &              \\
      & 0.037     &  0.977(23)  &  8.040(660) \\
0.2978 & 0.072    &  0.884(23)  &  5.875(434)  \\
 (351)& 0.105     &  0.779(25)  &  4.013(325)  \\
      & 0.136     &  0.700(42)  &  2.518(361)  \\
      & 0.166(1)  &  0.644(33)  &  2.208(222)  \\
      & 0.195(1)  &  0.582(42)  &  1.758(208)  \\
      & 0.249(1)  &  0.541(104) &  0.977(314)  \\
      & 0.274(2)  &  0.449(96)  &  0.792(271)  \\
      & 0.299(2)  &  0.408(116) &  0.791(330)  \\
      & 0.323(2)  &  0.326(91)  &  0.569(238)  \\
\hline
       & 0.0      &  1.156(47) &                \\
       & 0.037    &  0.967(31) &   6.579(980)   \\
0.2600 & 0.072    &  0.887(30) &   5.441(488)   \\
 (667) & 0.104    &  0.790(36) &   3.976(462)   \\
       & 0.135    &  0.628(46) &   2.425(453)   \\
       & 0.164(1) &  0.589(39) &   1.850(269)  \\
       & 0.192(1) &  0.507(42) &   1.510(255)   \\
       & 0.245(1) &  0.403(60) &   0.925(307)   \\
       & 0.270(1) &  0.335(62) &   0.638(250)  \\
\hline
\end{tabular}
\caption{Results on the axial nucleon form factors at $\beta=3.9$}
\label{tab:results 3.9}
\end{center}
\end{table}

\begin{table}[h]
\begin{center}
\begin{tabular}{c|c|c|c|}
\hline\hline
 $m_\pi$ (GeV)  & $(aQ)^2$  & $G_A$ & $G_p$   \\\hline 
 (no. confs)   &   &  &    \\\hline 
\multicolumn{4}{c}{$\beta=4.05$, $32^3\times 64$ }\\\hline
      &  0.0    &    1.173(24)  &             \\
      &  0.037  &    0.971(19)  & 8.896(485) \\
0.4653&  0.071  &    0.809(18)  & 5.063(269) \\
 (419)&  0.104  &    0.723(30)  & 3.735(269) \\
      &  0.134  &    0.625(51)  & 2.463(309) \\
      &  0.163  &    0.566(52)  & 2.064(229) \\
      &  0.191(1)&   0.481(83)  & 1.540(289) \\
      &  0.243(1)&   0.306(86)  & 0.879(283) \\
      &  0.268(1)&   0.181(173) & 0.310(319) \\
\hline
      &  0.0      &  1.175(31) &               \\
      &  0.037    &  0.961(29) &   8.213(645)  \\
0.4032&  0.071    &  0.842(32) &   5.206(388)  \\
 (326)&  0.103    &  0.792(56) &   4.027(429)  \\
      &  0.133    &  0.643(77) &   2.224(415)  \\
      &  0.161(1) &  0.522(50) &   1.499(215)  \\
      &  0.188(1) &  0.516(146)&   1.563(485)  \\
      &  0.238(1) &  0.209(50) &   0.584(188) \\
      &  0.262(1) &  0.168(73) &   0.420(204)  \\
\hline
      &  0.0      &   1.194(66) &                 \\
      &  0.037    &   0.873(46) &  7.165(1.089)  \\
0.2925&  0.070    &   0.735(47) &  4.273(516) \\
 (447)&  0.101    &   0.557(73) &  1.757(525) \\
      &  0.130(1) &   0.540(110)&  1.726(559) \\
      &  0.157(1) &   0.509(194)&  1.376(564) \\
      &  0.182(1) &   0.383(103)&  1.061(358) \\
\hline
\end{tabular}
\caption{Results on the axial nucleon form factors at $\beta=4.05$}
\label{tab:results 4.05}
\end{center}
\end{table}

\begin{table}[h]
\begin{center}
\begin{tabular}{c|c|c|c|}
\hline\hline
 $m_\pi$ (GeV)  & $(aQ)^2$  & $G_A$ & $G_p$    \\\hline
 (no. confs)   &   &  &     \\\hline  
\multicolumn{4}{c}{$\beta=4.2$, $32^3\times 64$ }\\\hline
      & 0.0      &   1.130(26) &             \\
      & 0.036    &   0.890(21) &  6.348(422)  \\
0.4698& 0.069    &   0.786(270)&  3.632(220) \\
(357) & 0.099    &   0.659(61) &  2.289(266) \\
      & 0.126    &   0.573(123)&  1.462(422) \\
      & 0.152(1) &   0.449(83) &  1.073(221) \\
      & 0.177(1) &   0.262(50) &  0.535(114) \\
      & 0.222(1) &   0.148(62) &  0.225(123) \\
\hline
\multicolumn{4}{c}{$\beta=4.2$, $48^3\times 96$ }\\\hline
       & 0.0        & 1.138(43) &                \\
       & 0.016      & 0.997(33) & 11.392(882)    \\
 0.2622& 0.032      & 0.856(23) &  5.860(473)   \\
  (245)& 0.046      & 0.759(28) &  3.845(417)   \\
       & 0.060      & 0.734(48) &  3.212(358)  \\
       & 0.072      & 0.634(30) &  2.295(224)  \\
       & 0.085      & 0.584(37) &  1.604(188)  \\
       & 0.108(1)   & 0.440(48) &  1.260(217) \\
       & 0.119(1)   & 0.414(39) &  0.946(151) \\
       & 0.129(1)   & 0.364(65) &  0.604(181) \\
       & 0.139(1)   & 0.328(76) &  0.452(201) \\
\hline
\end{tabular}
\caption{Results on the axial nucleon form factors at $\beta=4.2$}
\label{tab:results 4.2}
\end{center}
\end{table} 
\end{widetext}

\end{document}